\documentclass[conference]{IEEEtran}
\usepackage{cite}
\usepackage{amsmath,amssymb,amsfonts}
\usepackage{algorithmic}
\usepackage{graphicx}
\usepackage{textcomp}
\usepackage[dvipsnames]{xcolor}
\usepackage{siunitx}
\usepackage{tikz}
\usepackage{circuitikz}
\usepackage{pgfplots}
\pgfplotsset{compat=1.8}
\usetikzlibrary{decorations.markings}
\def\BibTeX{{\rm B\kern-.05em{\sc i\kern-.025em b}\kern-.08em
    T\kern-.1667em\lower.7ex\hbox{E}\kern-.125emX}}
\begin{document}

\title{Open Source Reconfigurable Intelligent Surface for the Frequency Range of 5 GHz WiFi
\thanks{This work was funded in the framework of BMBF MetaSec Project, Nr. 16KIS1236 by the German Ministry for Research and Education.}
}

\author{\IEEEauthorblockN{Markus Heinrichs}
\IEEEauthorblockA{\textit{High Frequency Laboratory} \\
\textit{TH Koeln - University of Appl. Sciences}\\
Cologne, Germany\\
markus.heinrichs@th-koeln.de}
\and
\IEEEauthorblockN{Aydin Sezgin}
\IEEEauthorblockA{\textit{Institute of Digital Communication Systems} \\
\textit{Ruhr University Bochum}\\
Bochum, Germany\\
aydin.sezgin@rub.de}
\and
\IEEEauthorblockN{Rainer Kronberger}
\IEEEauthorblockA{\textit{High Frequency Laboratory} \\
\textit{TH Koeln - University of Appl. Sciences}\\
Cologne, Germany\\
rainer.kronberger@th-koeln.de}
}

\maketitle

\begin{abstract}
Reconfigurable Intelligent Surfaces (RIS) have been identified as a potential ingredient to enhance the performance of contemporary wireless communication and sensing systems. Yet, most of the existing devices are either costly or not available for reproduction. To close this gap, a Reconfigurable Intelligent Surface for the frequency range of 5~GHz WiFi is presented in this work. We describe the designed unit cell, which is optimized for the full frequency range of 5.15 to 5.875~GHz. Standard FR4 substrate is used for cost optimization. The measured reflection coefficient of a rectangular RIS prototype with 256 elements is used for RF performance evaluation. Fabrication data and firmware source code are made open source, which makes RIS more available in real measurement setups. 
\end{abstract}

\begin{IEEEkeywords}
Reconfigurable Intelligent Surface, RIS, unit cell, WiFi, open source
\end{IEEEkeywords}

\section{Introduction}
Reconfigurable Intelligent Surfaces (RIS) are currently a major subject of research on wireless systems. Multiple potential application areas of RIS are being investigated around the world, from next generation  cellular networks (6G) through unmanned aerial vehicle assisted systems to physical layer security systems \cite{pan}\cite{li}. Nevertheless, the availability of RIS prototypes is still limited, which does not allow every research group to perform experiments in real measurement setups using RIS. In particular, the large amount of time required to develop a RIS can be a time-consuming and costly impediment. In order to facilitate the access to measurements with RIS, we have made our $\SI{5.5}{\giga\hertz}$ RIS design open source, similar to the OpenRIS for $\SI{3.5}{\giga\hertz}$ from \cite{rains}. The fabrication data is freely accessible and can be used without restriction, which enables everyone to build and utilize RIS. The frequency range of $5$ to $\SI{6}{\giga\hertz}$ is chosen since measurements at sub-$\SI{6}{\giga\hertz}$ frequencies are cost effective and widespread, not least because of their use in WiFi. The wavelength of approx. $\SI{55}{\milli\meter}$ at a center frequency of $\SI{5.5}{\giga\hertz}$ results in a size ratio that allows arrays with several hundred elements while at the same time being manufacturable with acceptable effort.

In the following, we describe our RIS unit cell and its principle of operation. Then we address the RF performance and present measurement results.

\section{Unit Cell}
\begin{figure}[!htb]
    \centering
    \begin{circuitikz}[scale=0.65, transform shape,
        text pos/.store in=\tpos,text pos=0.5,
        text anchor/.store in=\tanchor,text anchor={north:10pt},
        Tline/.style={
            draw,
            postaction={decorate,decoration={markings,mark=at position 10pt with {\coordinate (a) at (90:3.5pt);}}},
            postaction={decorate,decoration={markings,mark=at position \tpos with {\node at (\tanchor){\small #1};}}},
            postaction={decorate,decoration={markings,mark=at position \pgfdecoratedpathlength-10pt with {\coordinate (b) at (-90:3.5pt);\draw[fill=black!40](a) rectangle (b);}}}
        }]
        \centering
        \draw [draw=black, fill=SpringGreen] (0,1) rectangle (8,7.5); 
        \draw [draw=black, fill=Peach] (0.7,1.45) rectangle (7.2,7.05); 
        \draw [draw=black, fill=white, thick] (4,1.7) circle (0.1); 
        \draw [dashed, draw=black] (0.7,4.25) -- (1.1,4.25); 
        \draw [dashed, draw=black] (1.5,4.25) -- (7.2,4.25); 
        \draw [<->, draw=black, thick] (0,7.95) -- (8,7.95); 
        \node [align=center] at (4,8.15) {$w_{uc}=\SI{20.00}{\milli\meter}$};
        \draw [dashed, draw=black] (0, 7.6) -- (0, 8);
        \draw [dashed, draw=black] (8, 7.6) -- (8, 8);
        \draw [<->, draw=black, thick] (0.5,1) -- (0.5,7.5); 
        \node [align=center, rotate=90] at (0.3,4.25) {$l_{uc}=\SI{13.00}{\milli\meter}$};
        \draw [<->, draw=black, thick] (0.7,6.3) -- (7.2,6.3); 
        \node [align=center] at (4,6.5) {$w_p=\SI{17.21}{\milli\meter}$};
        \draw [<->, draw=black, thick] (1.4,1.45) -- (1.4,7.05); 
        \node [align=center, rotate=90] at (1.2,4.25) {$l_p=\SI{11.18}{\milli\meter}$};
        \draw [<->, draw=black, thick] (3.45,1.7) -- (3.45,4.25); 
        \node [align=center, rotate=90] at (3.25,2.975) {$o_v=\SI{5.46}{\milli\meter}$};
        \draw [dashed, draw=black] (3.4,1.7) -- (3.9,1.7);
        
        \draw [draw=Peach, fill=Peach] (1.5,0) rectangle (6.5,-0.05); 
        \draw [draw=SpringGreen, fill=SpringGreen] (1,-0.05) rectangle (7,-0.8); 
        \draw [draw=Peach, fill=Peach] (1,-0.6) rectangle (2,-0.65); 
        \draw [draw=Peach, fill=Peach] (3,-0.6) rectangle (7,-0.65); 
        \draw [draw=Peach, fill=Peach] (1,-0.8) rectangle (2,-0.85); 
        \draw [draw=Peach, fill=Peach] (2.2,-0.8) rectangle (2.8,-0.85); 
        \draw [draw=Peach, fill=Peach] (3,-0.8) rectangle (7,-0.85); 
        \draw [draw=Peach, fill=Peach] (2.4,-0.05) rectangle (2.6,-0.8); 

        \draw (2.5,-0.85) -- (2.5,-1.5)
            -- (1,-1.5) -- (1,-2.5);
        \draw [Tline] (1,-2.5) -- (4.5,-2.5);
        \draw (2.75,-2.1) node[align=center, anchor=center] {$TL_1$};
        \draw (2.75,-3.2) node[align=center, anchor=center] {$w=\SI{0.42}{\milli\meter}$\\$l=\SI{9.1}{\milli\meter}$};
        \draw (4.5,-2.5) node[spdt, anchor=in] (Sw) {}
            (Sw.in) node[left] {}
            (Sw.out 1) node[right] {}
            (Sw.out 2) node[right] {}
            to ++(1,0) node[ground] {};
        \draw [draw=black] (Sw)+(-0.5,-0.5) rectangle ++(0.5,0.5);
        \draw (Sw)+(0,0.5) node[above] {\it{RF-Switch}};
        \draw (Sw)+(0,-0.5) node[below] {BGS12P2L6};

    \end{circuitikz}
    \caption{Unit cell dimensions and schematic}
    \label{fig:unitCellAndSchematic}
\end{figure}
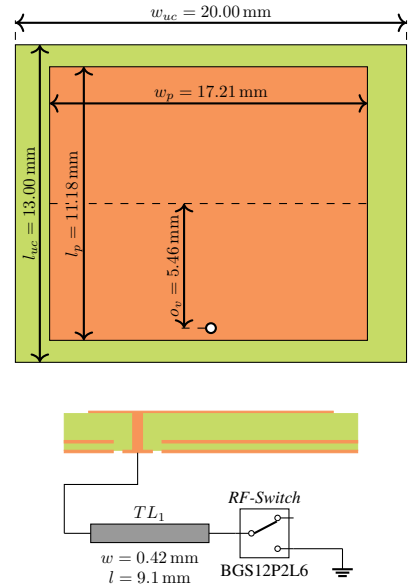

The RIS unit cell consists of a rectangular patch reflector above a ground plane. The patch on the front side is connected to the back side of the RIS by a plated through-hole connection (via), as in the case of a linearly polarized pin-fed patch antenna. We realized this by using a printed circuit board (PCB) with three copper layers, which can be manufactured externally in a standard process as a four-layered PCB with the second layer left empty. Thus, the unit cell can be implemented on a single PCB on which the components can be placed and soldered by machines, facilitating manufacturing. We used a low-cost FR4 substrate with a specified dielectric constant of $\epsilon_r=4.6$ and dissipation factor $\tan\delta=0.028$. A cross sectional view of the unit cell and its dimensions are depicted in Fig.~\ref{fig:unitCellAndSchematic}.

The shown circuit network is located on the back side of the unit cell and generates a binary switchable reflection coefficient. For this we use a BGS12P2L6 single pole, double throw (SPDT) RF switch in CMOS technology from Infineon Technologies \cite{infineon}, which switches the antenna port to either an open or short circuit. In this way, a phase shift of $\SI{180}{\degree}$ is achieved while keeping the DC power consumption of the RF switches at a low level.
The microstrip transmission line $TL_1$ is used to tune the RF performance. Our unit cell is optimized for the WiFi frequency range of $5.15$ to $\SI{5.875}{\giga\hertz}$. This means, that the magnitude of the reflection coefficient of the surface should be as large as possible in both switching states, with the propagation direction of the incident and reflected wave being directed orthogonal to the surface (normal incidence and reflection). The phase difference between the two switching states is optimized to be as close as possible to the ideal value of $\SI{180}{\degree}$ over the whole frequency range considered. By using these optimization goals, the resulting transmission line $TL_1$ has a width of $w=\SI{0.42}{\milli\meter}$ and a length of $l=\SI{9.1}{\milli\meter}$. Simulation and optimization of the unit cell is done in CST Microwave Studio using Floquet ports and frequency domain solver.

\begin{figure}[!htb]
    \centering
    \includegraphics[width=0.45\textwidth]{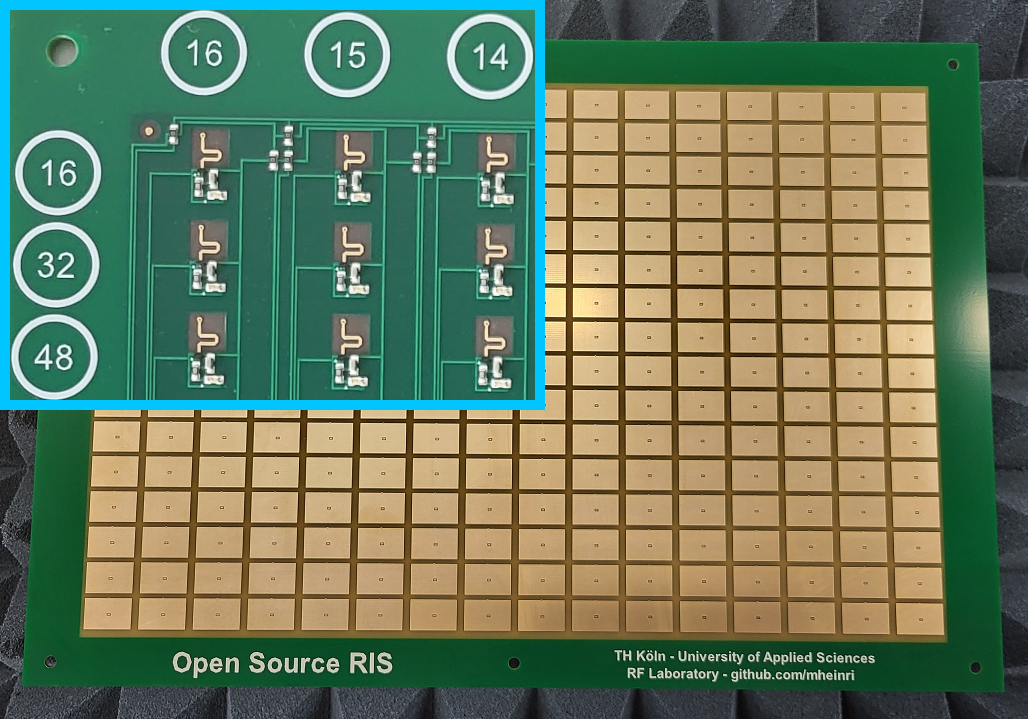}    
    \caption{RIS with 256 elements and sectional view of nine backside elements}
    \label{fig:pictureRIS}
\end{figure}

The fabricated RIS, which consists of $16 \times 16$ unit cells, is shown in Fig.~\ref{fig:pictureRIS}. A controller PCB is attached to the backside. This allows the RIS to be connected to a PC and controlled via a USB connection. Alternatively, it can be operated wirelessly with a battery and controlled via Bluetooth Low Energy (BLE).

\section{RF Measurements}
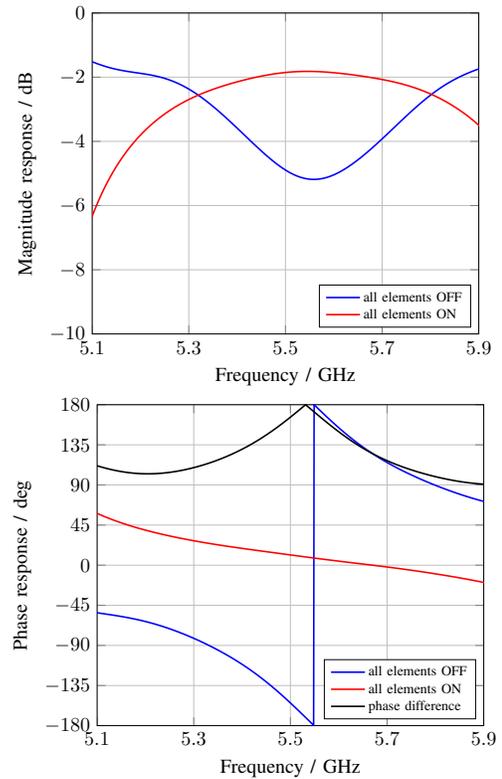
\begin{figure}[!htb]
    \centering
    \begin{tikzpicture}[scale=0.75, transform shape]
        \begin{axis}[
        xmin=5.1,
        xmax=5.9,
        ymin=-10,
        ymax=0,
        xtick={5.1,5.3,5.5,5.7,5.9},
        ytick={-10,-8,-6,-4,-2,0},
        xlabel={Frequency / GHz},
        ylabel={Magnitude response / dB},
        grid=both,
        legend style={nodes={scale=0.7, transform shape}, anchor=south east, at={(0.98,0.02)}},
        legend cell align=left
        ]
        \addplot [color=blue, mark=none, thick] table [x expr=\thisrowno{0}, y expr=\thisrowno{1}, col sep=comma] {RIS_data.csv};
        \addplot [color=red, mark=none, thick] table [x expr=\thisrowno{0}, y expr=\thisrowno{2}, col sep=comma] {RIS_data.csv};
        \addlegendentry{all elements OFF}
        \addlegendentry{all elements ON}
        \end{axis}
    \end{tikzpicture}
    \begin{tikzpicture}[scale=0.75, transform shape]
        \begin{axis}[
        xmin=5.1,
        xmax=5.9,
        ymin=-180,
        ymax=180,
        xtick={5.1,5.3,5.5,5.7,5.9},
        ytick={-180,-135,-90,-45,0,45,90,135,180},
        xlabel={Frequency / GHz},
        ylabel={Phase response / deg},
        grid=both,
        legend style={nodes={scale=0.7, transform shape}, anchor=south east, at={(0.98,0.02)}},
        legend cell align=left
        ]
        \addplot [color=blue, mark=none, thick] table [x expr=\thisrowno{0}, y expr=\thisrowno{3}, col sep=comma] {RIS_data.csv};
        \addplot [color=red, mark=none, thick] table [x expr=\thisrowno{0}, y expr=\thisrowno{4}, col sep=comma] {RIS_data.csv};
        \addplot [color=black, mark=none, thick] table [x expr=\thisrowno{0}, y expr=\thisrowno{5}, col sep=comma] {RIS_data.csv};
        \addlegendentry{all elements OFF}
        \addlegendentry{all elements ON}
        \addlegendentry{phase difference}
        \end{axis}
    \end{tikzpicture}
    \caption{Surface reflection coefficient (magnitude, phase and phase difference)}
    \label{fig:measuredS11}
\end{figure}

The reflection coefficient of the RIS is measured with a vector network analyzer while the RIS is located in the aperture of a horn antenna, which is specially designed for the measurement of surface reflection coefficients. The magnitude of the measured surface reflection coefficient is shown in Fig.~\ref{fig:measuredS11}~(top). The measurement is normalized to the reflection of a metal plate in the size of the RIS. The corresponding phase as well as the phase difference between the two states in which all elements are turned OFF or ON are shown in Fig.~\ref{fig:measuredS11}~(bottom). The phase difference is wrapped to values between $\SI{0}{\degree}$ and $\SI{180}{\degree}$.

In the WiFi frequency bands, the minimum surface reflection coefficient is $\SI{-5.2}{\decibel}$ at $\SI{5.56}{\giga\hertz}$ for the OFF state and $\SI{-4.8}{\decibel}$ at $\SI{5.15}{\giga\hertz}$ for the ON state. This could be improved slightly by using low-loss RF substrate, however, with a significant increase of material costs. The phase difference reaches its maximum of $\SI{180}{\degree}$ at $\SI{5.53}{\giga\hertz}$ and is minimal with a value of $\SI{92}{\degree}$ at $\SI{5.875}{\giga\hertz}$. It is under current investigation how this phase behaviour performs in practical systems.

\section{Conclusion}
We have presented an open source RIS for the frequency bands of $\SI{5}{\giga\hertz}$ WiFi. It offers a worse-case reflection coefficient of $\SI{-5.2}{\decibel}$ and a phase response between $\SI{92}{\degree}$ and $\SI{180}{\degree}$ over the whole frequency range considered. The RIS is low cost and easy to manufacture. It can be controlled easily and is enabled for wireless usage. Fabrication data, firmware source code and additional documentation for the presented open source RIS are accessible at GitHub \cite{heinrichs}.

\section{Acknowledgment}
This work was funded by the German Federal Ministry of Education and Research (BMBF) in the framework of Project MetaSec (Förderkennzeichen 16KIS1235 and 16KIS1236).

\end{document}